# Accelerating String Matching for Bio-computing Applications on Multi-core CPUs

Damayanthi Herath, Chathurika Lakmali and Roshan Ragel
Department of Computer Engineering
University of Peradeniya
Peradeniya, Sri Lanka

*Abstract*— Huge amount of data in the form of strings are being handled in *bio-computing applications* and searching algorithms are quite frequently used in them. Many methods utilizing on both software and hardware are being proposed to accelerate processing of such data. The typical hardware-based acceleration techniques either require special hardware such as general-purpose graphics processing units (GPGPUs) or need building a new hardware such as an FPGA based design. On the other hard, software-based acceleration techniques are easier since they only require some changes in the software code or the software architecture. Typical software-based techniques make use of computers connected over a network, also known as a network grid to accelerate the processing. In this paper, we test the hypothesis that multi-core architectures should provide better performance in this kind of computation, but still it would depend on the algorithm selected as well as the programming model being utilized. We present the acceleration of a string-searching algorithm on a multi-core CPU via a POSIX thread based implementation. Our implementation on an 8-core processor (that supports 16-threads) resulted in 9x throughput improvement compared to a single thread implementation.

## I. Introduction

String set matching has a vital importance in computational biology in places like DNA synthesis. An example is the PCR (Polymerase Gene Reaction) process in DNA synthesis [1], which amplifies a specific region of DNA to provide enough copies of that to be sequenced or tested. To use this process, biologists need to know the exact sequences that lie on either side of the region of interest. Hence finding the regularities in patterns is very useful for which string searching algorithms are used.

The definition for bio-computing can be given as the application of information technology and computer science to biological problems, in particular to issues that involve genetic sequences. Pattern matching questions in computational biology arise from the desire to know different characteristics about DNA sequences.

The new era is witnessing a remarkable increase in the discovery of the number of nucleotide and amino acid sequences and therefore the content of the biological databases seems to double frequently. Hence researches are being done in order to accelerate the algorithms used in computational biology such that they use minimal computer storage and would minimize the searching response time. The focus of such research is on hardware or software acceleration and also on accelerating through algorithm optimization.

As cited by Kostogryz in [2], the avalanche of data growth arises as a consequence of advances in the fields of molecular biology and genomics and proteomics. The challenge for biologist nowadays lies in the de-codification of this huge and complex data, in order to achieve a better understanding of how our genes shape and how our genome evolved, and therefore how we function and who we are.

Put in brief, string matching refers to locating occurrences of one or more strings/patterns within a larger string. There are many string matching algorithms available, which are useful in computational biology applications such as Smith-Waterman (SW) [14], Cloudburst [15] and Aho-Corasick [16]. These algorithms can be mainly divided into two categories [9]: (1) Bloom filters and (2) Exact string matching algorithms.

Bloom filters are probabilistic data structures that can be used to test whether a given element is within a set. On the other hand, exact string matching algorithms refer to the algorithms, which are based on finite state machines. Aho-Corasick, which is one of the most widely used algorithms for string searching because of its linear computational complexity and its ability to scale well while increasing the number of patterns, falls into the latter category. We have targeted the Aho-Corasick algorithm for the work described in this paper. Aho-Corasick is a multiple string-matching algorithm, which is capable of locating all occurrences of any of a finite number of keywords within a given string of text in a single pass.

To accelerate string-matching, many hardware-based approaches have been suggested in the past such as utilizing

FPGAs, GPU, Cell Broadband Engines (CBE) and General Purpose multi cores [3]. In addition, several approaches are being suggested to achieve improved performance (throughput) via software-based solutions. What we are focusing on in this paper is a multi-core CPU based software implementation for accelerating Aho-Corasick algorithm. We have implemented Aho-Corasick as POSIX threads and were able to achieve a 9x throughput improvement over a single thread implementation in an 8-core CPU that supports 16-threads.

The rest of the paper is organized as follows: In Section II we present related work and in Section III we discuss on the background details of our experiment focusing on Aho-Corasick algorithm and POSIX threads. Section IV and Section V presents methodology followed and results obtained respectively. In Section VI we conclude the paper.

## II. RELATED WORK

A large number of researches are being performed in order to find or derive efficient string matching algorithms compared to the existing ones as well as to find hardware/software accelerators for existing algorithms. Here we briefly present on evaluations done on different string matching algorithms and mainly focus on methodologies, which have been suggested to accelerate Aho-Corasick algorithm for string matching based on both hardware and software.

Eric Rouchka discusses various techniques and tools for solving various pattern matching questions in computational biology [4] and how these sequences can be modelled. String searching algorithms plays a vital role in these applications. There are many such algorithms suggested as was mentioned in the introduction. Benfano and Ning have presented a methodology [11] for evaluating string matching algorithms. They have developed an automatic simulation framework to evaluate the performance of different algorithms on multiprocessors.

Researches have been done in order to find means of optimizing the Aho-Corasick (AC) algorithm. For an example, in the Parallel Failure less-AC Algorithm (PFAC) [5] the authors are trying to overcome the problems in the direct implementation of Aho-Corasick algorithm on GPU and increase the throughput by increasing the parallelism. This research has been done focusing on string searching algorithms used in Network Intrusion Detection Systems (NIDS). There, they are removing the failure transitions so that there is no need to backtrack the state machine used, reducing the complexity of the algorithm. They have achieved this by allocating each byte of an input stream, a GPU thread to identify any virus pattern starting at the thread starting location. Their research was quite useful for us to understand the algorithm and the different methods we could develop to parallelize it.

Wei Lin and Bin Liu have presented a study on an improved Aho-Corasick algorithm, which is called P2-AC on SRAM [8]. The authors suggested a pipelined parallel approach on SRAMs. The study also provides comparison of several Aho-Corasick implementations and their performance measures.

In [12], a comparison on several software-based implementations of the Aho-Corasick algorithm for high performance systems has been presented. A detailed comparison has been presented on how each solution achieves the objectives of supporting large dictionaries, sustaining high performance, and enabling customization and flexibility using various data sets and considering shared-memory and distributed memory architectures.

Studies have been performed considering implementations of Aho-Corasick algorithm on FPGAs as well [6, 18]. It is shown that GPUs achieves comparable or higher speedups than CBE-based platforms for computation-intensive applications. On the flip side, the GPU-based solution spends a significant fraction of its time in CPU-GPU communication through PCI-X. In [3], the authors have analysed and concluded that General Purpose multi-core platforms provide the best overall speedup and also provide the maximum ease of porting of code. While the effort required for CBE is maximum, due to their reconfigurable nature, FPGA platforms provide the scope for implementing parallel architectures specifically optimized for certain applications. It is also shown that the hardware-based methods have shown a significant level of improvement over single CPU implementation. This study has been based on different algorithms.

Focusing on NIDS, a research has been done on high-speed string searching against large dictionaries on the Cell/B.E. processor [7]. They have parallelized Aho-Corasick on the IBM Cell/B.E. processor with the intention of performing exact string matching against large dictionaries. There they have focused on implementing the algorithm in C language using the CBE intrinsic language extension. It is suggested that the memory congestion plays a crucial role in determining the performance of Aho-Corasick algorithm and hence trying to optimize the implementation mainly based on three aspects as memory pressure, layout issues and hot spots.

A software based implementation of the Aho-Corasick algorithm on Cray XMT multithreaded shared memory machine was suggested in [10]. They were able to achieve scalable high performance, independent of the input stream or the pattern set being analyzed by making use of features in XMT architecture and algorithmic strategies.

In this paper, we are proposing a software-based method for improving the performance of the Aho-Corasick algorithm using a multi-core processor by utilising its inherent support for implementing parallel threads. We developed the parallel pattern matching machine using POSIX threads and evaluated its throughput improvement and scalability.

III. BACKGROUND

The approach taken was to design a suitable representation of Aho-Corasick algorithm for a multiprocessor system, to implement it, and to analyse its performance (throughput) and scalability on different multiprocessor configurations. In this section, we will present how Aho-Corasick algorithm works in brief and how POSIX threads are used in general for implementing parallel applications.

*A. The Aho-Corasick Algorithm*

Aho-Corasick [16] is a popular string-matching algorithm that is simple and capable of finding a finite set of key words within a given input string in a single pass. Mainly the algorithm consists of two parts and they are:
  1. Constructing a string-matching machine from the given keywords, and
  2. Processing the input string in a single pass using the string-matching machine.

In the rest of this subsection, we give an example illustrating how Aho-Corasick algorithm functions. Fig. 1 depicts the finite state machine constructed for matching the keywords *HIS, SHE and HERS* following the first stage of the algorithm.

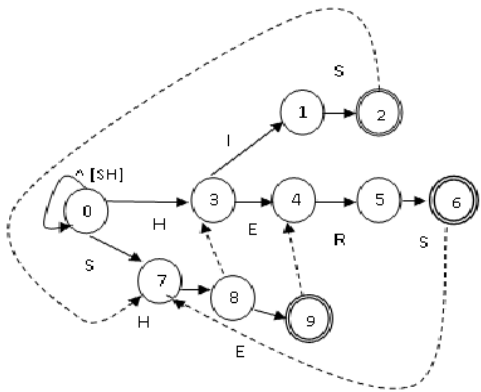

Figure 1.  Pattern matching machine developed for identifying the patterns HIS, SHE, and HERS

Solid lines in Fig. 1 represent valid transitions and dotted lines represent the failure transitions. Failure transitions are used for backtracking patterns starting at different locations. Here, failure transitions that lead back to state 0 are not shown for clarity.

In the second stage of Aho-Corasick, when an input character (from the input string that need to be matched) and the current state (of the finite state machine) are given, the machine checks for a valid transition and if it fails, it transits to the state pointed by the failure transition. Then the machine reads the same input character until it causes a valid transition.

For an example, consider the input string "SHERS": the machine has to recognize two patterns "SHE" and "HERS" in the input string as shown in Fig. 2. It is worth noting that "SHE" and "HERS" are overlapping in the input string and Aho-Corasick matches/identifies overlapping patterns as well.

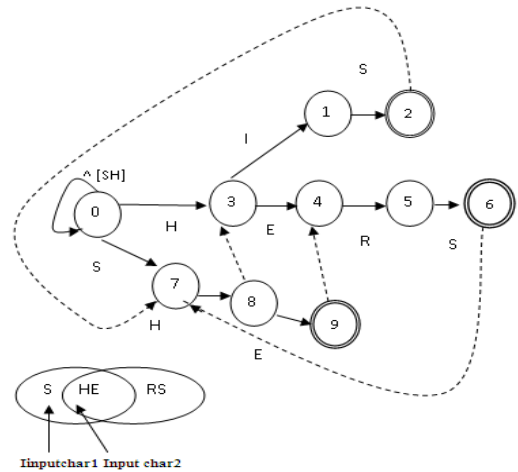

Figure 2.  Recognition of the patterns in the input string "SHERS"

Now, in stage two of Aho-Corasick, the finite state machine initially being at state 0, it transits from state 0 to state 7 (for input character "S"), then state 7 to state 8 (for the input character "H") and then state 8 to state 9 (for input character "E") identifying the pattern "SHE". The next input character "R" does not have a valid transition at state 9 and hence the next state transition is to state 4 directed by failure transition and then reading the same input "R" it does a valid transition from state 4 to state 5. Eventually the machine reaches state 6, which is the final state of the pattern "HERS". Therefore, the finite state machine has identified two patterns, "SHE" and "HERS" in the input string "SHERS".

*B. POSIX Threads*

The definition of a thread can be put as an independent flow of control inside an address space. Threads can be used to implement parallelism in shared memory multiprocessor architectures. POSIX threads or Pthreads refers to the C language threads programming interface for UNIX standardized by IEEE POSIX 1003.1c standard which enables to create a new thread within the caller process. Another quite widely used method of implementing parallel processing in UNIX is "fork" which will create a new process, which becomes the child process of the caller.

Experimental results have shown that *Pthread* utility can provide better results than a fork mainly because a thread can be created with much less operating system overheads than a fork as a fork generates a separate process execution [13].

## IV. DESIGN AND IMPLEMENTATION

In this section, we brief on our parallel processor design and present our implementation.

### A. Thread Assignment Methodology

At the beginning of our design, we explored the possibility of parallel execution through two different architectures:
1. Splitting the input into separate chunks and processing them for the string matching in separate threads.
2. Splitting the pattern file and developing different pattern machines in separate threads and passing the input to each machine.

When we consider typical bio-computing applications, the pattern text using which we are searching strings in the input text, would be quite large and hence the time taken for developing the pattern matching machine would considerably effect the total time taken. Hence we took the second approach listed above in this experiment where we tried to optimize the total time based on time taken for building the pattern matching machine. As shown in Fig. 3, the large pattern file (where the patterns are stored one per line) was divided into separate chunks and each thread developed a pattern matching machine (a separate finite automaton) based on these chunks and the same input file was processed individually by each machine.

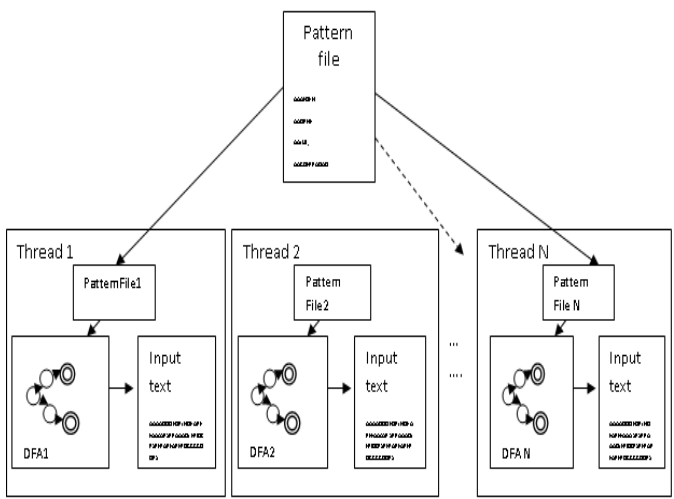

Figure 3. The thread based implementation to achieve better throughput by separately manipulating the pattern files (assuming N number of threads)

### B. Implementation Detail

First task was to find a suitable implementation of Aho-Corasick algorithm. We could find several libraries available that implement the Aho-Corasick algorithm written in languages like Java, C, C++, Python and Haskell. We used a module called "*MultiFast-v0.6.2*"[1]. The reason for choosing *Multifast* was that it provides an easy to understand code base and a neat implementation. It contains a basic implementation, which would take an array of strings as a set of finite pattern strings and a string against which the patterns would be matched, and then outputs the details on the pattern matched and its position in the input string.

Additionally *Multifast* provides the facility of analyzing data reads from files. Both the set of patterns and the input string can be given from files. We mainly utilized this functionality in our work. The approach we took was to achieve parallelism via threads. We used C programming language and p_thread library for thread handling.

First, the pattern file was split into a given number of threads and then the same numbers of threads are created; each pattern chunk was passed onto a separate thread. A given thread took a pattern and the input file and searched for the given patterns in the input. Then the average time taken for each execution was measured by performing the same experiment many times. The throughput was measured as number of patterns matched per second and this process was repeated varying the number of threads being used. This setup was run in different types of processors. All the tests were carried out on GNU/Linux platform.

## V. EXPERIMENTAL EVALUATION

We analysed the results mainly considering the following aspects:
1. Throughput: The effect of utilizing multi-core processors, which are capable of running threads.
2. Scalability: The effect of changing the input file size (the string size) used in our implementation.

$$Throughput = \frac{\text{Total number of patterns matched}}{\text{The average time taken}} \quad (1)$$

Different processors we used differed mainly with respect to the optimum number of parallel threads that they supported (Table I). Considering the fact 1 listed above, it was tested how the support for parallelism provided by each processor can affect the throughput, against running the implementation as a single process (on the same processor). Here, the throughput was calculated using Equation 1.

---

[1] C library for Aho-Corasick algorithm freely available from Sourceforge.net (http://sourceforge.net/projects/multifast/)

For considering the scalability of our approach, the experiments were performend by increasing the input file size from 6MB to 30MB in 6MB increments. If the throughput variation did not significantly change with the increasing file size, it would assure us that more time was spent on creating the pattern matching machine than searching for patterns in the input file and therefore the scalability of the suggested method. Fig. 4 shows the variation of throughput for different number of threads for each processor listed in Table I. From Fig. 4, we could observe that using the suggested thread based implementation we could utilise the parallelism supported by processors and achieve improved performance. We could see higher throughput values for multi-core CPUs supporting more optimal number of threads. For an example, in Fig. 4 consider the variation for Intel dual core T4500 processor. The processor supports a maximum of two parallel threads and for a 6MB size input file, we could obtain 1.7x times throughput improvement compared to single thread implementation on the same processor.

TABLE I. PROCESSOR CONFIGURATIONS USED

| Processor ID | Description | # of cores | Optimum # of threads supported |
|---|---|---|---|
| P1 | Intel Pentium4 | 1 | 1 |
| P2 | Pentium Dual Core T4500 | 2 | 2 |
| P3 | Intel Core i3 370M | 2 | 4 |
| P4 | Intel® Xeon® CPU X7560 | 8 | 16 |

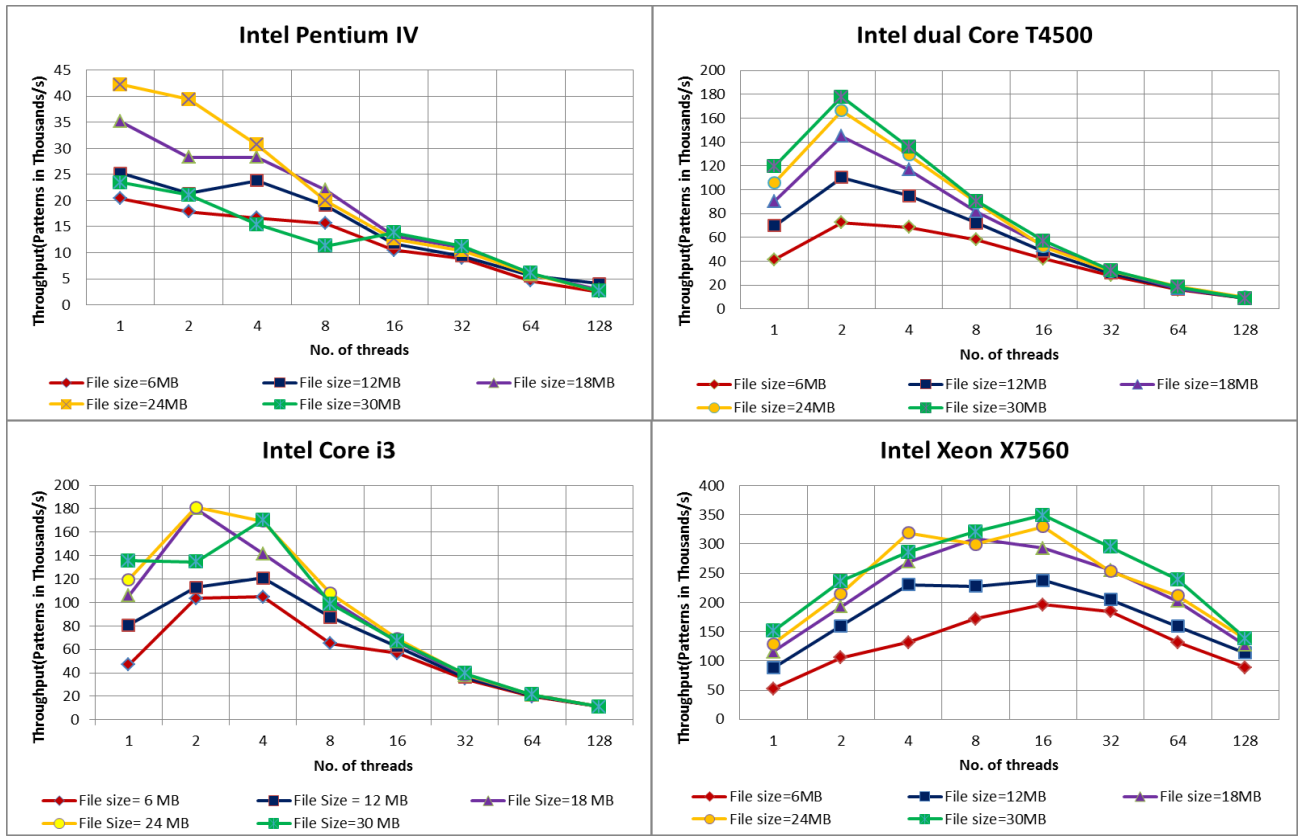

Figure 4. Variation of throughput with number of threads for different file sizes in different processors

Fig. 4 also depicts the variation of throughput for different number of threads as the input file size was varied. Here, we see that for a given processor, the behaviour remains unchanged though the input file size is varied. That is, we still observe that utilising the optimal number of threads would result in the maximum throughput. It implies that we could obtain a significant improvement by optimising the time taken for developing the pattern matching machine via separately processing the pattern file in different threads.

From Fig. 4 we can also see that for a given processor and a given number of threads, throughput has increased as the input file size was increased. It is because the average time remained constant as input file was made larger and still the number of patterns identified increased hence increasing the throughput. It also ensures our basis in this design, i.e. the time taken for developing the pattern machine would be more significant than that for searching patterns and hence varying the input file size would not affect the execution time significantly.

Still, in Fig. 4 we observe that when the number of threads is large, the throughputs for different file sizes remain closer to each other in contrast to what we presented in the last paragraph. The analysis in [17] explains this scenario. The paper [17] presents that for a given multi-core processor performance varies linearly until it saturates available number of threads and after that point, execution time would increase without giving any performance gain. In our case too, the execution time has increased, as the number of threads was made larger than optimal number of parallel threads supported by the processor. Since number of patterns too was increased the ratio remained closer to each other.

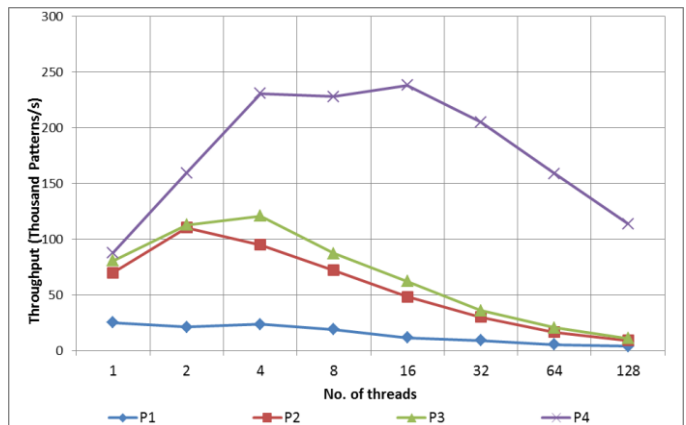

Figure 5. Variation of throughput with number of threads for different processor types for an input file of 12 MB

Fig. 5 illustrates a comparison between different processors. It has been plotted considering an input file of size 12MB and we observe that our implementation is capable of utilising the optimum number of parallel threads supported by each processor and also the fact that the suggested thread based implementation on a multi-core processor could provide better throughput than that of single threaded implementation. We could obtain 9 times throughput improvement in the Intel® Xeon® X7560 (supporting optimum 16 threads) processor compared to that of the Pentium 4 processor.

## VI. CONCLUSION

Given the increasing need for matching large patters, new techniques are being suggested to improve the performance of the pattern matching algorithms being used. In this paper, we have presented a methodology to achieve improved performance of Aho-Corasick pattern matching algorithm on a multi-core CPU through parallel manipulation of pattern files using POSIX thread utility. Results obtained on different processors confirm the usefulness and scalability of our methodology. Future work on this would include optimising the methodology to utilize hardware that supports better parallelism such as GPGPU.